# Cavity-enhanced dual-comb spectroscopy

Birgitta Bernhardt [1], Akira Ozawa [1], Patrick Jacquet [2], Marion Jacquey [2], Yohei Kobayashi [3], Thomas Udem [1], Ronald Holzwarth [1,4], Guy Guelachvili [2], Theodor W. Hänsch [1,5], Nathalie Picqué [1,2]

1. Max Planck Institut für Quantenoptik, Hans-Kopfermann-Str. 1, 85748 Garching, Germany
2. Laboratoire de Photophysique Moléculaire, CNRS, Bâtiment 350, Université Paris-Sud, 91405 Orsay, France
3. The Institute for Solid State Physics, University of Tokyo, 5-1-5 Kashiwanoha, Kashiwa, Chiba 277-8581 Japan
4. Menlo Systems GmbH, Am Klopferspitz 19, 82152 Martinsried, Germany
5. Ludwig-Maximilians-Universität München, Fakultät für Physik, Schellingstrasse 4/III, 80799 München, Germany

**The sensitivity of molecular fingerprinting is dramatically improved when placing the absorbing sample in a high-finesse optical cavity, thanks to the large increase of the effective path-length. As demonstrated recently, when the equidistant lines from a laser frequency comb are simultaneously injected into the cavity over a large spectral range, multiple trace-gases may be identified within a few milliseconds. Analyzing efficiently the light transmitted through the cavity however still remains challenging. Here, a novel approach, cavity-enhanced frequency comb Fourier transform spectroscopy, fully overcomes this difficulty and measures ultrasensitive, broad-bandwidth, high-resolution spectra within a few tens of µs. It could be implemented from the Terahertz to the ultraviolet regions without any need for detector arrays. We recorded, within 18 µs, spectra of the 1.0 µm overtone bands of ammonia spanning 20 nm with 4.5 GHz resolution and a noise-equivalent-absorption at one-second-averaging per spectral element of $3 \cdot 10^{-12}$ cm$^{-1}$Hz$^{-1/2}$, thus opening a route to time-resolved spectroscopy of rapidly-evolving single-events.**

Cavity-enhanced and cavity-ring-down spectroscopies[1,2] are widely used for ultrasensitive spectroscopic absorption measurements and they have led to remarkable progress in fundamental spectroscopy and non-intrusive trace-gas sensing. While these techniques were initially mostly practiced with tunable narrow bandwidth lasers, dramatic advances[3-8] have been achieved with the coherent coupling of a laser frequency comb (FC) to a high finesse-cavity containing the sample. The spectral analysis of the light transmitted through the cavity is performed with dispersive spectrometers, usually equipped with detector arrays. This resulted[3] in massively parallel spectra recorded in a spectral span as broad as 15 nm with 25 GHz resolution, 1.4 ms acquisition time and a minimum-detectable-absorption coefficient $\alpha_{min}$ of $6.3 \cdot 10^{-7}$ cm$^{-1}$. Subsequent refinements in this promising experimental approach led to spectral resolution up to 800 MHz[5,9], $\alpha_{min}$ coefficient improving to $8 \cdot 10^{-10}$ cm$^{-1}$ within 30s measurement time[6] and have already enabled practical applications to trace gas detection[6,7]. These schemes share the drawback of using dispersive spectrometers. They limit the resolution obtainable in a motionless short measurement, even though sweeping the comb parameters[7] or implementing Vernier techniques[8] proved successful in improving the resolution, at the price of longer and sequential recordings. Additionally, large detector arrays are not conveniently available in the mid-infrared molecular fingerprint spectral region, where most molecules have intense rovibrational signatures. Frequency combs have prompted the alternative method of FC Fourier transform spectroscopy[10-20] (FTS) which does not encounter such spectral bandwidth or resolution limitations while providing extremely short measurement times. For instance, spectra[13] spanning 120 nm in the region of 1.5 µm are measured within a recording time of 42 µs, and 5 GHz-resolution. However, this method presents sensitivities that are several orders of magnitude too low for the various applications linked to trace-gas detection.

Here we present an approach which fully overcomes this dichotomy. A proof-of-principle experiment combines, without trade-off, the ultra-high sensitivity of cavity-enhancement and the broad spectral bandwidth, high resolution, high accuracy, very fast acquisition times of FC-FTS, by the multiplex (i.e. using a single photodetector) analysis of the modes of a FC simultaneously injected in a high-finesse resonator.

An optical FC[21,22] typically provides, in a single laser beam, several hundred thousands phase-coherent optical frequency markers with very narrow linewidths. Interferences between two independent combs, with slightly different repetition frequencies, can benefit optical diagnostics and precision spectroscopy, by taking advantage of motionless novel Fourier transform spectroscopy[10-20]. The beat notes between pairs of lines from the two combs occur in the radio-frequency domain thus providing a down-converted image of the optical spectrum. In the simplest approach the sample is probed by only one comb and the encoded spectral information is observed by heterodyne detection with the second comb, acting as a reference. Simultaneous and accurate access to a broad spectral bandwidth is provided within a short measurement time. This can physically be equally understood in terms of time-domain interferences, multi-heterodyne detection, linear optical sampling or cross-correlation between two electric fields.

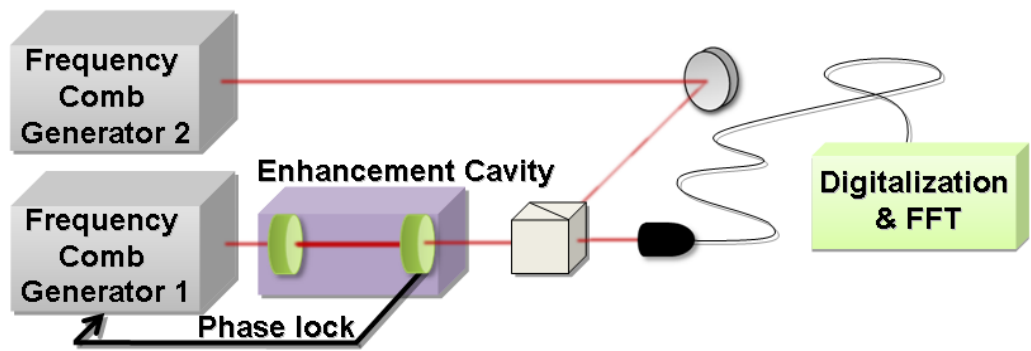

*Figure 1. Experimental set-up.* Two frequency comb generators, named 1 and 2, have slightly different line spacing. Frequency comb 1 is transmitted through the resonant high-finesse cavity, which holds the sample under study. The repetition frequency of frequency comb 1 is phase-locked onto the cavity free spectral range. The light transmitted through the cavity is heterodyned against the comb 2 on a single fast photodetector, yielding a down-converted radio-frequency comb containing information on the ultrasensitive absorption losses experienced by each line of the comb 1. The electrical signal is digitized and is Fourier-transformed using a fast Fourier transform (FFT) algorithm.

In our experimental set-up (Fig. 1), the pulses from the interrogating 1040 nm Ytterbium-doped fiber comb, named 1, are amplified with an Ytterbium-doped fiber amplifier and mode-matched into a 230-cm long resonant high-finesse ring cavity placed in a vacuum-tight chamber, which contains the sample. The cavity has a free spectral range of 130 MHz, which matches the comb repetition frequency. The cavity mirrors provide a finesse $F > 1200$ and a group-delay dispersion $< 31$ fs$^2$ for 20 nm of spectral bandwidth between 1030 and 1050 nm. The effective interaction length between the light field and the sample is therefore dramatically enlarged to 880 m, as the absorption enhancement factor is indeed $F/\pi$ in a ring resonator. The comb is locked to the cavity with a Pound-Drever-Hall scheme[23]. The light transmitted through the cavity is recombined using a fiber coupler with the reference comb 2, which is free-running. The difference between the two comb repetition frequencies has been chosen between 200 and 600 Hz. The two combs beat on a fast photodiode and the electrical signal is filtered with a low pass filter and digitized with a high-resolution acquisition board. The time-domain interference signal (Fig.2) is Fourier-transformed to reveal the absorption spectrum. Additional experimental details may be found in the supplementary material.

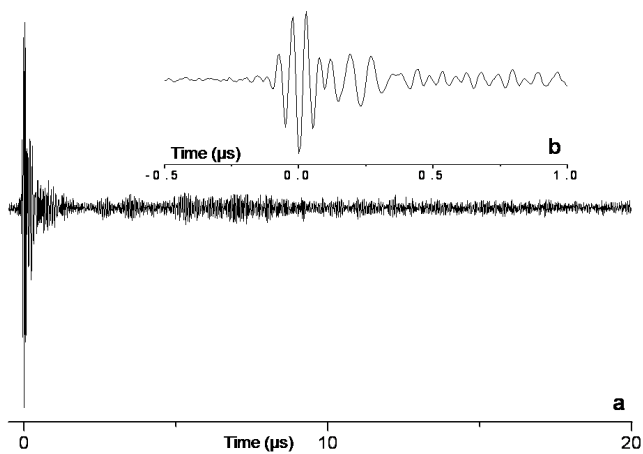

*Figure 2. Time-domain interferogram.* a) An interferogram of acetylene acquired within 20 µs, without averaging, is displayed. This unweighted interferogram leads to a spectrum with 4.5 GHz unapodized resolution: when an interferogram is unweighted, the shape of the spectral line is the convolution of the true spectrum and a sinc function (i.e., the Fourier transform of the boxcar finite-measurement time truncation function). Instead, if one used a well-chosen weighting numerical function, the true spectrum would be convolved with the Fourier transform of this function. This operation is called apodization, as it considerably reduces the amplitude of the sidelobes of the convolving function at the expense of a loss in resolution. The interferogram displayed in a) repeats itself at a period, which is the inverse of the difference in the repetition frequencies of the two combs. The burst, arbitrarily set at 0 µs corresponds to the overlap of two femtosecond pulses. b) Zoom on the burst area. Apart from the burst, i.e. for times longer than 1 µs, the interferometric signal exhibits the typical modulation due to the molecular lines. It only occurs on one side of the burst, as the absorbing sample held in the high-finesse resonator just interacts with one of the two combs.

The 1.0-µm region is the seat of weak molecular overtone bands that can most often hardly be detected in standard laboratory conditions. The electrical and mechanical anharmonicities allow overtones and combination transitions to occur, even though their intensity dramatically drops off with increasing number of simultaneously excited normal vibrations. Extensive knowledge of these excited rovibrational levels proves however crucial for the accurate description of the anharmonicity of molecular Hamiltonians and the understanding of astronomical and atmospheric observations.

In this proof-of-principle experiment, spectra of acetylene and ammonia have been recorded. The region of the $3\nu_3$ band[24] of acetylene has already been studied in particular due to its relatively high line-strengths and to its usefulness for frequency metrology. An absorption spectrum (Fig.3), i.e. the Fourier-transform of a single time-domain interferogram sequence without averaging, resolves the rovibrational lines with a good signal-to-noise ratio (SNR). The spectral span extends from 1025 nm to 1050 nm. The unapodized resolution is 4.5 GHz. By evaluating the root-mean-square absorption noise level at positions where no signal is detected, the SNR in the spectral domain for the most intense line is about 100, and the recording time T = 23 µs for M = 1500 spectral elements (SE) (M = span/resolution). The minimum-detectable-absorption coefficient $\alpha_{min}$ is of the order of 8 $10^{-8}$ cm$^{-1}$. To account for FC-FTS multiplex nature, the noise-equivalent-absorption coefficient (NEA) at 1s-time-averaging per SE is defined as $\alpha_{min}(T/M)^{1/2}$. Its value is 8 $10^{-12}$ cm$^{-1}$Hz$^{-1/2}$ per SE.

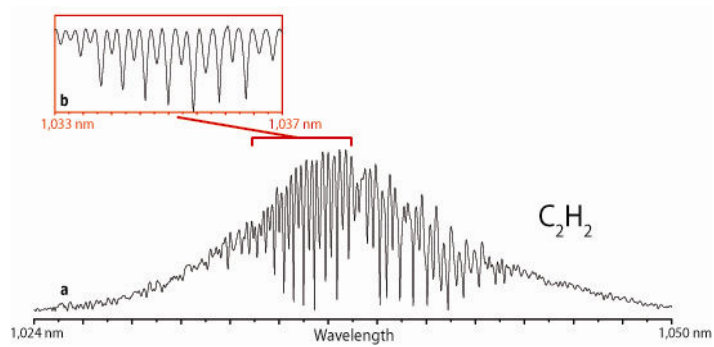

*Figure 3. Cavity-enhanced FC-FT spectrum of acetylene. The overtone bands of $C_2H_2$ recorded according to the cavity-enhanced FC-FTS principle illustrated in Figure 1 are plotted with a linear intensity scale. The high-finesse cavity is filled with 3 hPa of acetylene in natural abundance. The laser spectrum in a) extends from 1025 to 1050 nm. The absorption spectrum reports the acetylene intensity alternation of the $3\nu_3$ vibrational band centered at 1037.4 nm. Signatures seen in b) around 1035 nm belong to the R-branch (from R(19) to R(1)).*

To match the region of the ammonia absorption lines, the center of the cavity transmission spectrum is shifted to 1045 nm by acting on the grating in front of the photodiode used in the Pound-Drever-Hall detection scheme. This results (Fig.4) also in a better SNR, culminating at 380. To our knowledge, the $3\nu_1$ band of $NH_3$ is rotationally resolved for the first time, while the need for such spectral data has been broadly recognized[25,26] in particular for the radiative transfer modeling of the atmosphere of Jovian planets. The ammonia molecule is an oblate symmetric rotor, which can rapidly (~$10^{-11}$ s) invert, leading to two equilibrium positions for the N atom at the two sides of the $H_3$ plane. The facile interconversion by tunnelling of the inversion doubling causes an energy-level pattern for each form of $NH_3$ which together with the additional effects of resonances, make the overtone spectrum of ammonia irregular and crowded. Revealing its rotational fine structure is consequently critical for its exhaustive elucidation, as already demonstrated for the fundamental transitions. In Fig. 4, the cavity transmission spans about 20 nm and the spectrum with 4.5 GHz resolution, is measured within 18 µs. The minimum-detectable-absorption coefficient $\alpha_{min}$ and NEA at 1s-time averaging per SE are 2 $10^{-9}$ cm$^{-1}$ and 3 $10^{-12}$ cm$^{-1}$Hz$^{-1/2}$ per SE, respectively. Our proof-of-principle experiment already demonstrates, with a 100-fold shorter measurement time, a $\alpha_{min}$ coefficient, which is 300-fold better than the one reported in Ref.3.

Our promising experimental concept can be further improved. The spectral bandwidth is presently limited by the cavity mirrors, but the multiplex spectrometer principle allows for the measurement of multi-octave spanning spectra. Special mirror designs managing dispersion may match[3] the cavity modes and the comb components across 100 nm simultaneously. Such a bandwidth can easily be achieved by the spectral broadening of the combs with nonlinear optical fibers. The resolution can also be further increased so that individual comb lines are resolved[11,14]. Scanning the comb and interleaving[11] successive spectra can provide a resolution that is ultimately only limited by the width of the comb lines. For applications to trace-gas detection, reaching the mid-infrared region is a crucial objective, as the strength of the fundamental molecular lines drastically enhances the detection sensitivity. The $\alpha_{min}$ coefficient of $2 \cdot 10^{-9}$ cm$^{-1}$ achieved here, in 18 µs recording time, would for instance grant a minimum-detectable-concentration of 3 parts-per-trillion (ppt) and 210 ppt of $^{12}C^{16}O_2$ at 4.2 µm and 2.7 µm, respectively. This would improve to 0.1 parts-per-trillion and 10 parts-per-trillion with Doppler-limited resolution in ~550 µs recording time, respectively. Although FC oscillators are not yet directly available in this region, non-linear frequency conversion processes, already demonstrated[27] with an optical parametric oscillator spanning simultaneously up to 300 nm in the 2.8-4.8 µm range, provide efficient comb sources. FC-FTS only needs one detector, easily available in practically all spectral regions. With this additional advantage, it can be envisioned that cavity-enhanced FC-FTS will assume a position of dominance for the measurement of real-time ultrasensitive spectra in the molecular fingerprint region, similarly to the one that Michelson-based FTS holds for long for broadband Doppler-limited accurate spectra. For time-resolved applications, the interferogram periodicity at the $1/(f_{rep1}-f_{rep2})$ rate can be exploited. Furthermore, this acquisition rate may be increased[18] by varying the repetition frequency of one of the combs. Consequently, time-resolved sequences of broadband spectra reporting the evolution of a source every tens of µs could be measured, opening intriguing potential for the real-time monitoring of dynamic single-events.

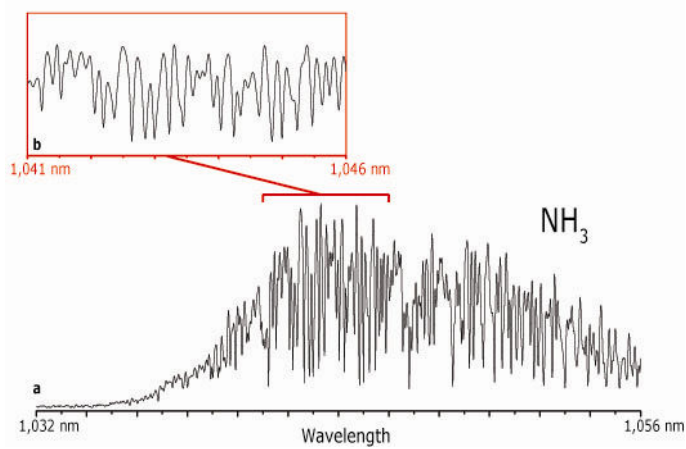

*Figure 4. Cavity-enhanced spectrum of the crowded region of the $3\nu_1$ overtone band of ammonia.* The cavity is filled with 50 hPa of ammonia. The weak transitions are observed at high resolution for the first time to our knowledge and are of interest for the modeling of radiative transfer in the atmospheres of the giant Jovian planets.

**Acknowledgments** Research conducted in the scope of the European Associated Laboratory "European Laboratory for Frequency Comb Spectroscopy". Support has been provided by the Max Planck Foundation and, for the PhD fellowship of P.J., the Délégation Générale de l'Armement. The expert help of Diana Höfling and Tobias Wilken in the Yb lasers operation is warmly acknowledged.

# Supplementary Material
# Cavity-enhanced dual-comb spectroscopy

Birgitta Bernhardt [1], Akira Ozawa [1], Patrick Jacquet [2], Marion Jacquey [2], Yohei Kobayashi [3], Thomas Udem [1], Ronald Holzwarth [1,4], Guy Guelachvili [2], Theodor W. Hänsch [1,5], Nathalie Picqué [1,2]

1. Max Planck Institut für Quantenoptik, Hans-Kopfermann-Str. 1, 85748 Garching, Germany
2. Laboratoire de Photophysique Moléculaire, CNRS, Bâtiment 350, Université Paris-Sud, 91405 Orsay, France
3. The Institute for Solid State Physics, University of Tokyo, 5-1-5 Kashiwanoha, Kashiwa, Chiba 277-8581 Japan
4. Menlo Systems GmbH, Am Klopferspitz 19, 82152 Martinsried, Germany
5. Ludwig-Maximilians-Universität München, Fakultät für Physik, Schellingstrasse 4/III, 80799 München, Germany

**Frequency comb Fourier transform spectroscopy principle**

Frequency comb Fourier transform spectroscopy has already been discussed in several publications [S1-S9] and we only recall here its general principle for the clarity of the present letter.

The spectrum of a frequency comb consists of a comb of laser modes and can be described by the well-known equation [S10]:

$$f_{n,i} = f_{0,i} + n f_{rep,i}$$

where n is a large integer number ($\sim 10^5$), i addresses the laser 1 or 2 in our experiment, $f_{0,i}$ is the carrier-envelope offset frequency that is induced by the difference in group and phase velocities of the laser pulses and $f_{rep,i}$ is the repetition frequency of the laser i.

The beating signal $I$ between the combs 1 and 2 is detected by a fast photodiode and may be written as:

$$I(t) = \sum_{n}' A_n \cos\left((f_{01} - f_{02} + n(f_{rep1} - f_{rep2}))t\right)$$

$A_n$ is the product of the amplitude of the electric fields of the lasers, also involving the amplification by the Yb amplifier, the enhancement by the cavity and possible attenuation induced by gas absorption inside the cavity in the first of the two beating arms. Analogously to the use of a Michelson interferometer, the optical frequencies $f_{n,i} = f_{0,i} + n f_{rep,i}$ are down-converted to $f_{0,1} - f_{0,2} + n (f_{rep,1} - f_{rep,2})$. This down-converted spectrum lies in the radio-frequency domain between 0 and $f_{rep,i}/2$. This signal is Fourier-transformed to reveal the spectrum.

**Detailed experimental set-up for cavity-enhanced frequency comb Fourier transform spectroscopy with Ytterbium fiber lasers**

In the present experiment, two femtosecond Ytterbium fiber lasers with slightly different repetition frequencies ($\Delta f = f_{rep,1} - f_{rep,2} \sim 200\text{-}600$ Hz) are used as spectrometric devices. One of these lasers is coherently coupled to a resonant high-finesse cavity which contains the absorbing sample.

Our experimental set-up is shown in detail in Figure S1. The output of the first Ytterbium doped fiber laser (Menlosystems "Orange" prototype, repetition frequency ~ 130 MHz, average power: 100 mW, pulse length τ ~ 2.2 ps which may be compressed down to 100 fs) is amplified in an Ytterbium fiber amplifier after passing a stretcher fiber. The stretcher fiber (Sumitomo, length = 3m) broadens the pulses to ~ 15 ps while it pre-compensates the third order dispersion that is introduced by the amplifier fiber and the subsequent compressor gratings. The stretched pulses are sent into a 3.2 m long Ytterbium-doped, polarization-maintaining double-clad fiber (core diameter: 20 µm). This amplifier fiber is reversely pumped by a diode laser running at a wavelength of 976 nm (60 A, 75 W). The amplified pulses are compressed to a duration of about 100 fs via a transmission grating pair. The amplifier can reach more than 17 W of average output power behind the compressor. This exceeds the requirements of the present experiment. Here, an output power of about 1 W has been sufficient, and the amplifier was actually included due to the different original purpose this experiment was designed for i.e. producing a frequency comb in the extreme ultra-violet region (XUV) via intra-cavity high harmonic generation (HHG) in a noble gas [S11-S14]. In the present experiment, the amplified comb light is sent through a ring enhancement cavity that is filled with the gas of interest. Inside the enhancement cavity, the pulses are circulating for a certain life time that is dependent of the cavity's Finesse $F$. Due to this fact, the interaction length of the light with the gas is increased by a factor of $F/\pi$. The losses due to the gas absorption are enhanced resulting in an increased sensitivity.

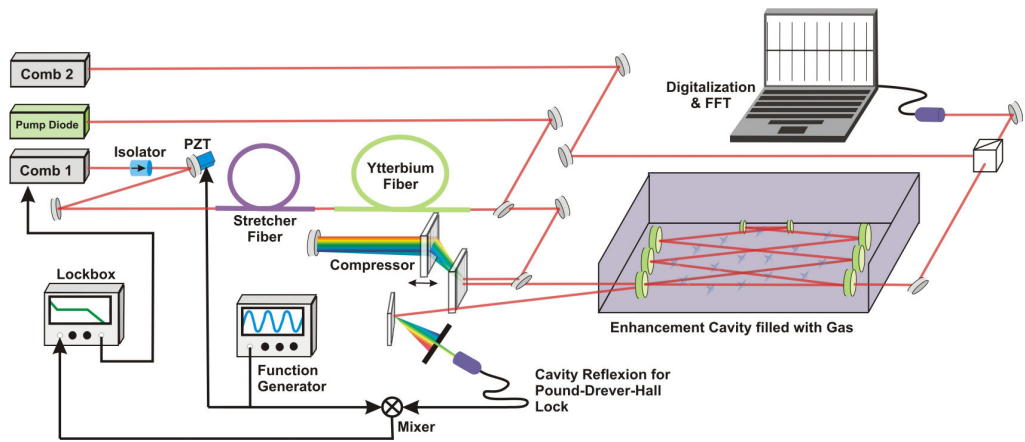

**Figure S1. Detailed experimental set-up for cavity-enhanced frequency comb Fourier transform spectroscopy with Yb fiber lasers.**
*The pulses from the interrogating comb 1 are amplified and coupled into a resonant high-finesse cavity, which is filled with the absorbing gas sample. To keep the enhancement cavity in resonance with the interrogating comb, the comb repetition frequency is locked to the cavity free spectral range with a Pound-Drever-Hall scheme. The light leaking outside the cavity beats with the reference comb 2 on a fast photodiode and the electric signal is digitized with a high resolution acquisition board. The absorption spectrum is computed with a fast Fourier transform algorithm.*

To achieve the coherent addition of the intra-cavity pulses, the cavity free spectral range has to match with the laser repetition frequency. In other words, the cavity round-trip time T has to be the reciprocal value of the laser's repetition frequency $f_{rep,1} = 1/T$. In practice, this coupling is implemented via a Pound-Drever-Hall lock [S15]. For this purpose, the light that is reflected by the cavity input coupler is detected after it is diffracted by a grating and filtered by a slit for wavelength selection of the locking point. The use of the grating results also in a better signal-to-noise ratio of the error signal. This error signal evolves while sidebands of the laser repetition frequency are generated via a piezoelectric transducer (PZT) on the laser output beam (mirror on blue box in Fig. S1). The PZT modulates at 668 kHz the phase of the laser beam and the light reflected by the cavity is compared with the modulation signal produced by a function generator (for further explanations see [S15,S16]). The comparing mixer extracts the part that is at the same frequency as the modulation signal and generates an error signal. A proportional-integral-controller feeds back the error signal to a piezoelectric transducer with 45 kHz bandwidth inside the laser resonator and maintains the system onto resonance. The carrier-envelope offset frequency $f_{0,1}$ of the comb is tuned manually by tilting an intra-laser-cavity wedge to improve the overlap of the comb frequencies with the cavity modes.

As can be seen in Figure S1 (light violet box on the right), our enhancement cavity consists of eight mirrors: 6 plane and 2 concave mirrors (radius of curvature = 38 mm) producing a tight focus between the latter two mirrors. This rather complex cavity setup is also due to the initial XUV generation experiment the cavity was designed for. In practice, a simpler cavity with only two mirrors would produce similar results for cavity-enhanced frequency comb Fourier transform spectroscopy. Six of the cavity mirrors have a high reflectivity coating with a reflectivity value of R ~ 99.98 %. The output coupler of R ~ 99.86% determinates the reflectivity of the input coupler to R ~ 99.74 % via impedance matching. This results in a cavity finesse of F ~ 1200.

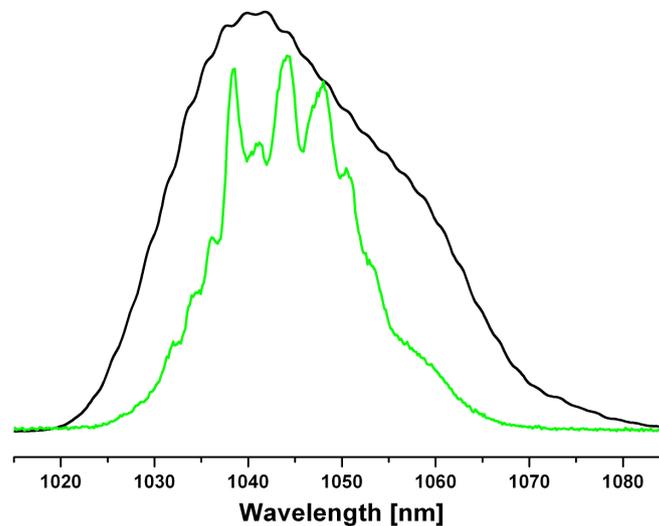

*Figure S2: The spectrum of the Yb-amplifier frequency comb (black), smoothed over 50 points, incident on the optical cavity. The spectrum transmitted from the cavity when the comb frequencies are locked to the cavity modes (green).*
The cavity filtering is due to dispersion inside the cavity (introduced by the mirrors and the gas inside the cavity) and to non-optimum locking conditions.

Figure S2 shows the spectrum of the incident Yb-amplifier frequency comb and the spectrum transmitted from the high-finesse cavity when the comb frequencies are locked to the cavity modes. Both spectra are measured with a low-resolution grating spectrometer. More than 30 nm of the spectrum are efficiently coupled into the cavity with a filtering of the incident comb spectrum due to frequency mismatch of the comb frequencies and cavity modes on the wings of the amplifier spectrum. Optimizing continuously the cavity transmission by locking the comb to the cavity proved crucial in the present experiment, as intensity noise is known to be the dominant noise source limiting the overall performance of a Fourier transform spectrometer. The continuous matching of the comb to the cavity is very sensitive to acoustic and vibration-induced noise, which gets converted to intensity noise on the cavity transmission, and therefore on the interferogram, if the feedback loop does not have sufficient bandwidth. Further improvements in our system therefore involve increasing the bandwidth of servo-control loop on the comb repetition frequency and adding active control on the carrier-envelope offset frequency.

For frequency comb Fourier transform spectroscopy measurements, the light that is transmitted through the output coupler of the resonant cavity is then overlapped with the second Ytterbium fiber laser (an in-house developed model, P = 100 mW, uncompressed pulse length τ ~ 1.5 ps) with a slightly different repetition frequency. Acoustic insulation is achieved by placing the laser assembly in a wood-compound enclosure with high air sound-absorption. The beat signal is generated by a fiber coupler with the ratio of 90:10 (90% of cavity transmission, 10 % of second fiber laser, in figure 1 displayed as a beam splitter cube) for a matched power balance and finally detected by a fast photodiode. After a low pass filter of 70 MHz that provides non-redundant information, the signal is amplified and digitized by a high resolution digitizer on a personal computer [S1]. Home-made programs compute and display the spectra.